\documentclass[pdflatex,sn-mathphys-num]{sn-jnl}

\usepackage{graphicx}%
\usepackage{multirow}%
\usepackage{amsmath,amssymb,amsfonts}%
\usepackage{amsthm}%
\usepackage{mathrsfs}%
\usepackage[title]{appendix}%
\usepackage{xcolor}%
\usepackage{textcomp}%

\usepackage{booktabs}%
\usepackage{algorithm}%
\usepackage{algorithmicx}%
\usepackage{algpseudocode}%
\usepackage{listings}%

\theoremstyle{thmstyleone}%
\theoremstyle{thmstyletwo}%

\theoremstyle{thmstylethree}%

\begin{document}

\title[Higher genus polylogarithms associated with 
the D'Hoker-Hidding-Schlotterer connection]
{Higher genus polylogarithms associated with 
the D'Hoker-Hidding-Schlotterer connection}


\author{\fnm{Takashi} \sur{Ichikawa}}\email{ichikawn@cc.saga-u.ac.jp}




\affil{\orgdiv{Department of Mathematics}, 
\orgname{Faculty of Science and Engineering}, 
\\
\orgaddress{\street{Saga 840-8502}, \country{Japan}}}





\abstract{
We study higher genus polylogarithms associated with the D'Hoker-Hidding-Schlotterer (DHS) connection  defined on families of Riemann surfaces of any genus with marked points. 
Under separating degenerations of Riemann surfaces, 
we show that the DHS connection can be constructed by gluing the Brown-Levin (BL) connections 
on elliptic curves and the Knizhnik-Zamolodchikov (KZ) connections on pointed Riemann spheres. 
Therefore, 
the associated higher genus polylogarithms can be described explicitly as power series 
in deformation parameters and their logarithms for the families of Riemann surfaces 
whose coefficients are expressed by multiple zeta values and elliptic multiple zeta values. 
}




\maketitle

\section{Introduction}

Polylogarithms on families of pointed Riemann surfaces are central objects 
in calculating Feynman integrals and string amplitudes. 
Based on remarkable results 
(see \cite{GiEG, BMW, BB, BMMS, BDDT, BDDPT1, BDDPT2, BSZ1, BDDPT3, BK, MS1, MS2, W} 
and there in) on this subjects in the genus zero and one cases, 
many researchers containing Baune, Broedel, D'Hoker, Enriquez, Hidding, Im, Lisitsyn, Schlotterer 
and Zerbini recently studied higher genus polylogarithms and concerning kernel functions. 
Especially, D'Hoker-Hidding-Schlotterer \cite{DHS} constructed higher genus polylogarithms 
as iterated integrals of a single-valued and modular invariant, 
but non-meromorphic connection with regular singularities at punctures on a Riemann surface 
which is called the DHS connection. 
This connection is a natural higher genus extension of the Brown-Levin (BL) connection 
on an elliptic curve \cite{BL} which was already applied to calculating one-loop Feynman integrals 
and string amplitudes (cf. \cite{BMMS, BDDT, BDDPT1, BDDPT2, BSZ1, BDDPT3, BK, MS1, MS2}), 
and it has rich mathematical structure since the description is given by 
Arakelov Green functions and modular tensors. 
Therefore, in order to give theoretical and practical contributions to particle physics, 
it is hoped to obtain explicit formulas of higher genus polylogarithms associated with the DHS connection. 
We note that D'Hoker-Enriquez-Schlotterer-Zerbini \cite{DESZ} related the DHS connection with 
the Enriquez connection \cite{E1} (cf. \cite{BBILZ} for its Schottky-Kronecker form) which is 
a meromorphic multi-valued connection with regular singularities 
at punctures on a given Riemann surfaces. 
See also \cite{BBILM, DS1, DS2, DS3, DS4} for the the Fay identities and the multi-variable version 
of these connections. 
Furthermore, Enriquez-Zerbini \cite{EZ1, EZ2, EZ3} constructed 
a holomorphic single-valued connection with irregular singularities at punctures. 

The aim of the present paper is to combine the result of \cite{DESZ} with results 
of \cite{F, I1, I2, I3, I4} on variations of Abelian differentials and the Enriquez connection 
for the description of the DHS connection and the associated DHS higher genus polylogarithms. 
By studying the variation of the DHS connection 
under separating degenerations of pointed Riemann surfaces, 
we show that this connection can be constructed by gluing the BL connections on elliptic curves 
and the Knizhnik-Zamolodchikov (KZ) connections on pointed Riemann spheres. 
Therefore, we can calculate the DHS polylogarithms 
on local families of pointed Riemann surfaces as products of monodromies 
of the DHS connections in the case of genus $\leq 1$.  
Furthermore, by the analytic continuation based on a higher genus extension 
of the cluster configuration structure in the genus $0$ case \cite{Br, AHHL, L}, 
we obtain explicit formulas of the DHS polylogarithms on degenerating families 
of pointed Riemann surfaces. 
The formulas are expressed as power series in deformation parameters (and their logarithms) 
associated with the families whose coefficients are expressed by 
multiple zeta values and their elliptic version. 
We note that our results are useful even in computing the DHS polylogarithms for genus one curves 
with multiple marked points.

\section{The Enriquez connection and its variation} 

\subsection{The Enriquez connection}

We consider a compact Riemann surface $R$ of any genus $g$, 
and denote its universal covering space by $\tilde{R}$ with the associated projection 
$\pi : \tilde{R} \rightarrow R$. 
Let $\{ A_{i}, B_{i} \}_{1 \leq i \leq g}$ be a basis of the first homology cycles which is symplectic, 
namely, their intersections satisfy that $A_{i} \cdot A_{j} = B_{i} \cdot B_{j} = 0$ 
and $A_{i} \cdot B_{j}$ is the Kronecker delta $\delta_{ij}$. 
Then there exists a unique basis $\{ \omega_{i} \}_{1 \leq i \leq g}$ 
of the space of holomorphic Abelian differentials on $R$ which is normalized for $\{ A_{i} \}_{i}$ 
in the sense that $\oint_{A_{i}} \omega_{j} = \delta_{ij}$, 
and hence one can define the period matrix $\Omega = (\Omega_{ij})_{i,j}$ 
of the surface $R$ with $\{ A_{i}, B_{i} \}_{i}$ by 
$$
\oint_{B_{i}} \omega_{j} = \Omega_{ij}. 
$$ 
Choosing the cycles $A_{i}$ and $B_{i}$ so that they share a common base point $q$, 
we regard these cycles as generators of the fundamental group $\pi_{1}(R; q)$ 
of $R$ with base point $q$ satisfying 
$$
[A_{1}, B_{1}]  [A_{2}, B_{2}] \cdots [A_{g}, B_{g}] = 1, 
$$
where $[A_{i}, B_{i}]$ denotes the commutator $A_{i} B_{i} A_{i}^{-1} B_{i}^{-1}$. 
Then one can take a fundamental domain $\mathcal{D}$ in $\tilde{R}$ for the action of 
$\pi_{1}(R; q)$ which is obtained by cutting $R$ along these cycles. 

Following \cite{DESZ, DS1}, 
we review results on the Enriquez connection which was given by Enriquez \cite{E1} 
in terms of a meromorphic connection form 
$$ 
\mathcal{K}_{R} (x, y; a, b) = \mathcal{K}^{\rm E}_{R} (x, y; a, b) 
$$ 
on $\tilde{R}$ with only simple poles 
which is valued in the freely completed Lie algebra $\mathfrak{g}$ with generators 
$a_{i}, b_{i}$ $(i = 1,..., g)$ corresponding to $A_{i}, B_{i}$ respectively. 
We denote its coefficient functions by  
$$
g^{i_{1} \cdots i_{r}}_{j}(x, y) = (g_{R})^{i_{1} \cdots i_{r}}_{j}(x, y) 
\quad (r \geq 0, \ i_{1},..., i_{r} \in \{ 1,..., g \})
$$
which are related to those used in \cite{E1} by 
$g^{i_{1} \cdots i_{r}}_{j}(x, y) = (-2 \pi \sqrt{-1})^{r} \omega^{i_{1} \cdots i_{r}}_{j}(x, y)$. 
These coefficient functions, also referred to as Enriquez kernels, 
are uniquely defined by the following properties. 
The Enriquez kernel $g^{i_{1} \cdots i_{r}}_{j}(x, y)$ is a $(1, 0)$-form in $x \in \tilde{R}$ 
and a scalar in $y \in \tilde{R}$ which is meromorphic for $x, y \in \tilde{R}$ 
and locally holomorphic in the complex moduli of pointed Riemann surfaces. 
In the above fundamental domain $\mathcal{D}$, 
$g^{i_{1} \cdots i_{r}}_{j}(x, y)$ is holomorphic in $x, y \in \mathcal{D}$ for $r \geq 2$, 
it can have a simple pole in $x$ at $y$ such as 
$g^{i}_{j}(x, y) - \delta_{ij} dx/(x - y)$ is holomorphic for $r = 1$, 
and is given by $g^{\emptyset}_{j}(x, y) = \omega_{j}(x)$ for $r = 0$. 
The monodromies in $x$ of $g^{i_{1} \cdots i_{r}}_{j}(x, y)$ around the $A$-cycles are trivial,
and those around the cycle $B_{l}$ are given by 
$$
g^{i_{1} \cdots i_{r}}_{j}(B_{l} \cdot x, y) = 
g^{i_{1} \cdots i_{r}}_{j}(x, y) + \sum_{k=1}^{r} \frac{(-2 \pi \sqrt{-1})^{k}}{k!} 
\delta^{i_{1} \cdots i_{k}}_{l} g^{i_{k+1} \cdots i_{r}}_{j}(x, y), 
$$
where $B_{l} \cdot x$ denotes the action of the element $B_{l} \in \pi_{1}(R, q)$ 
on the point $x \in R$, 
and the generalized Kronecker symbol is defined as 
$\delta^{i_{1} \cdots i_{k}}_{l} = \delta_{i_{1} l} \cdots \delta_{i_{k} l}$. 
Furthermore, by \cite{DESZ, DS1}, 
these forms satisfy the following conditions. 

\begin{itemize} 

\item 
trivial $A$ monodromies in $y$, 
and $B$ monodromies given by 
$$
g^{i_{1} \cdots i_{r}}_{j}(x, B_{l} \cdot y) = 
g^{i_{1} \cdots i_{r}}_{j}(x, y) + 
\delta^{i_{r}}_{j} \sum_{k=1}^{r} \frac{(2 \pi \sqrt{-1})^{k}}{k!} 
g^{i_{1} \cdots i_{r-k}}_{j}(x, y) \delta^{i_{r-k+1} \cdots i_{r-1}}_{l}. 
$$

\item 
the periods around $A$-cycles on the boundary of the fundamental domain $\mathcal{D}$ 
or any $y$ in the interior of $\mathcal{D}$ are given in terms of Bernoulli numbers $B_{r}$ as 
$$
\oint_{A_{k}} g^{i_{1} \cdots i_{r}}_{j}(t, y) = 
(-2 \pi \sqrt{-1})^{r} \frac{B_{r}}{r!} \delta^{i_{1} \cdots i_{r} k}_{j}. 
$$

\end{itemize}
Then the $\mathfrak{g}$-valued Enriquez connection $d - \mathcal{K}_{R}$ is expressed as 
$$
\mathcal{K}_{R}(x, y; a, b) = \sum_{r = 0}^{\infty} \sum_{1 \leq i_{1},..., i_{r}, j \leq g} 
g^{i_{1} \cdots i_{r}}_{j} (x, y) \ [b_{i_{1}}, [b_{i_{2}}, \cdots [b_{i_{r}}, a_{j}] \cdots ]]. 
\eqno(2.1) 
$$ 

In the genus one case, 
the Enriquez connection was shown to be derived from 
the elliptic Knizhnik-Zamolodchikov-Bernard (KZB) connection \cite{CEE} 
(cf. \cite[Section 8]{E1} and \cite[Section 4.2]{DS1}), 
and we give this explicit formula following \cite[Section 2.3]{DESZ}. 
See \cite{E2, E3, H} for the arithmetic nature of this connection and the associated polylogarithms. 
Put $\tau = \Omega_{11}$ as the restriction of the period matrix $\Omega$ to $g = 1$, 
and denote by $E$ the associated elliptic curve $\mathbb{C}/(\mathbb{Z} + \tau \mathbb{Z})$ 
with ${\rm Im}(\tau) > 0$. 
Then the Enriquez connection $\mathcal{K}_{E}$ satisfies the following explicit formula 
in terms of the odd Jacobi theta function: 
$$
\vartheta_{1}(x) = 2 q^{1/8} \sin(\pi x) 
\prod_{n=1}^{\infty} (1 - q^{n}) (1 - e^{2 \pi \sqrt{-1} x} q^{n}) (1 - e^{-2 \pi \sqrt{-1} x} q^{n}), 
$$
where $q = e^{2 \pi \sqrt{-1} \tau}$. 
We define the Kronecker function $F(x; \alpha)$ (also known as Kronecker-Eisenstein series) by 
$$
F(x; \alpha) = \frac{\vartheta'_{1}(0) \, \vartheta_{1}(x + \alpha)}{\vartheta_{1}(x) \, \vartheta_{1}(\alpha)}
$$
which is meromorphic in $x \in \mathbb{C}$ and regarded as multi-valued function on $E$ satisfying 
$$
F(x + 1; \alpha) = F(x; \alpha), \quad F(x + \tau; \alpha) = e^{-2 \pi \sqrt{-1} \alpha} F(x; \alpha). 
$$
Then the associated integration kernels $g^{(r)}$ are defined as 
$$
F(x; \alpha) = \alpha^{-1} + \sum_{r=1}^{\infty} g^{(r)}(x) \alpha^{r-1}, 
$$
and the elliptic Enriquez connection form $\mathcal{K}_{E}$ is given by  
\begin{align*}
\mathcal{K}_{E}(x, p; a, b) 
& = dx \, F(x - p; B) Ba; \ BX := [b, X] 
\\
& = dx \left( a + \sum_{r=1}^{\infty} g^{(r)}(x-p) B^{r} a \right). 
\end{align*}
Denote by $\mathbb{CP}^{1}/(0 = \infty)$ the singular complex curve obtained from 
the Riemann sphere $\mathbb{CP}^{1}$ with coordinate $z$ by identifying $z = 0$ and $\infty$. 
Then the map $x \mapsto e^{2 \pi \sqrt{-1} x}$ gives an identification of 
$E = \mathbb{C}/(\mathbb{Z} + \tau \mathbb{Z})$ with the deformation of 
$\mathbb{CP}^{1}/(0 = \infty)$ by the relation $\xi_{0} \cdot \xi_{\infty} = q$, 
where $\xi_{0} = z$ and $\xi_{\infty} = 1/z$ are local coordinates at $0$ and $\infty$ 
respectively. 
Therefore, 
$E$ degenerates to $\mathbb{CP}^{1}/(0 = \infty)$ under ${\rm Im}(\tau) \rightarrow \infty$, 
and then it is shown in \cite[Section 12]{H} that $\mathcal{K}_{E}$ degenerates to 
the KZ connection form on $\mathbb{CP}^{1}$ given by 
$$
\frac{dz}{z-1} [b, a] + \frac{dz}{z} \left( \frac{B}{e^{B} - 1} \right) a. 
\eqno(2.2)
$$

\subsection{Variation of the Enriquez connection} 

For each $i = 1,..., g$, 
let $(E_{i}; t_{i})$ be an elliptic (complex) curve, i.e., a Riemann surface of genus one 
with one marked point $t_{i}$, 
and take a symplectic homology basis $\{ A_{i}, B_{i} \}$ of $E_{i} \setminus \{ t_{i} \}$. 
Denote by $R_{0}$ the singular complex curve obtained from 
the Riemann sphere $\mathbb{CP}^{1}$ with $g$ marked points $x_{1},..., x_{g}$ 
and the elliptic curves $(E_{1}; t_{1}),..., (E_{g}; t_{g})$ by identifying $x_{i}$ and $t_{i}$ $(i = 1,..., g)$. 
For complex parameters $s_{i}$ with small $|s_{i}|$, 
put $s = (s_{1},...,s_{g})$, 
and denote by $\{ R_{s} \}_{s}$ the family of deformations of $R_{0}$ defined as 
$$
\xi_{i} \cdot \theta_{i} = s_{i} \quad (i = 1,..., g) 
\eqno(2.3)
$$
for local coordinates $\xi_{i}$ (resp. $\theta_{i}$) at $x_{i}$ (resp. $t_{i}$) such that 
$\xi_{i}(x_{i}) = 0$ (resp. $\theta_{i}(t_{i}) = 0$). 
Then one can take neighborhoods $U_{i}$ of $x_{i} \in \mathbb{CP}^{1}$ and 
$V_{i}$ of $t_{i} \in E_{i}$ $(i = 1,..., g)$ such that $R_{s}$ is obtained from $R_{0}$ by identifying 
the boundaries $\partial U_{i}, \partial V_{i}$ of $U_{i}, V_{i}$ respectively via (2.3). 
In other words, 
$R_{s}$ is obtained by gluing around of $\partial U_{i}$ and $\partial V_{i}$ via (2.3) such that 
the inside (resp. outside) of $\partial U_{i}$ are mapped to the outside (resp. inside) of $\partial V_{i}$. 
Therefore, if $s \rightarrow {\bf 0}:= (0,...,0)$, 
then (2.3) become the local equations $\xi_{i} \cdot \theta_{i} = 0$ 
at the singular points $x_{i} = t_{i}$ on $R_{0}$ which are called nodal (see {\bf Figure 1}). 
\vspace{-5ex}

\hspace{-7ex} \includegraphics[width=150mm]{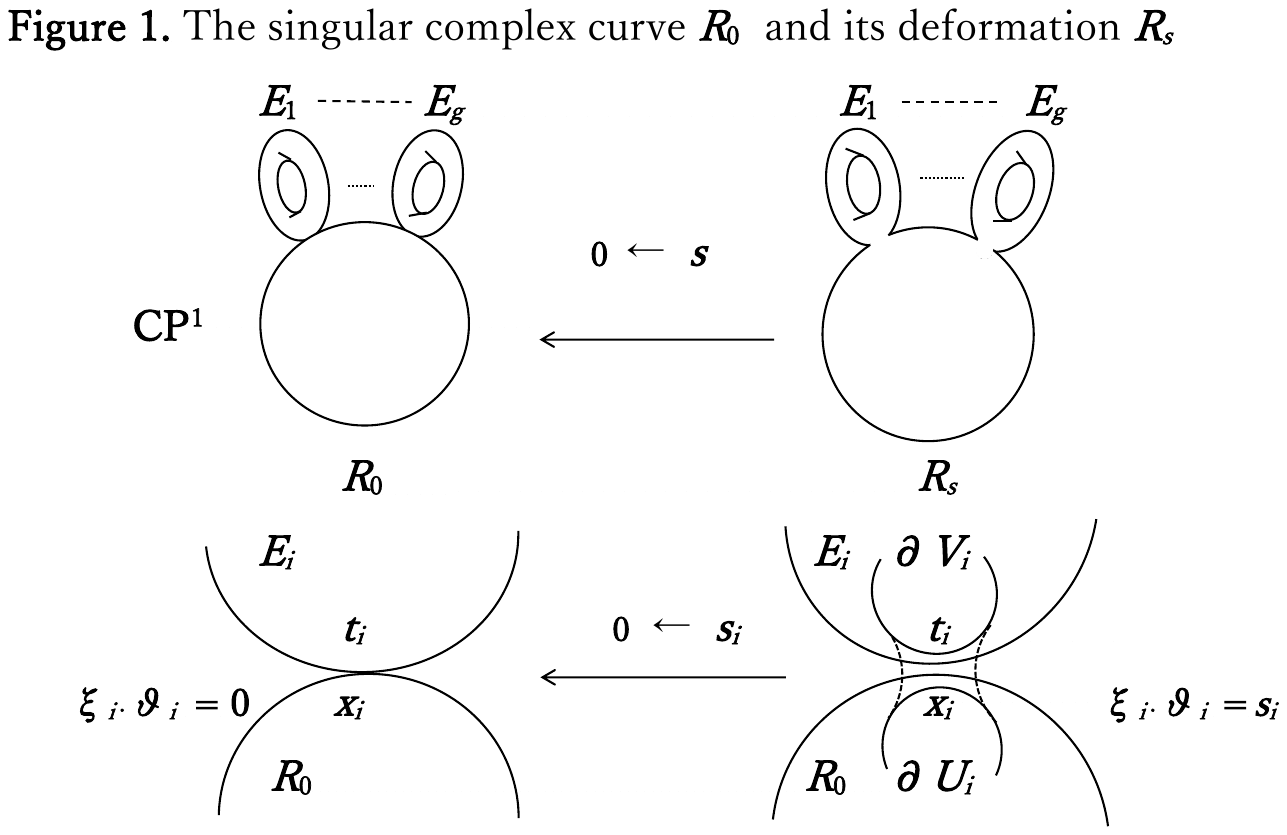}%
\vspace{-7ex}

\noindent
The above $\{ A_{i}, B_{i} \}_{1 \leq i \leq g}$ 
gives a symplectic homology basis of $R_{s}$, 
and points on $E_{i} \setminus \{ t_{i} \}$ and on $\mathbb{CP}^{1} \setminus \{ x_{1},..., x_{g} \}$ 
are identified with those on $R_{s}$ if $s$ is sufficiently small. 

For each $i = 1,..., g$, 
there exists a unique family $(\omega_{s})_{i}$ of holomorphic Abelian differentials 
on the above $R_{s}$ with $s_{1},..., s_{g} \neq 0$ which is normalized for $\{ A_{i} \}_{i}$, 
i.e., $\oint_{A_{i}} (\omega_{s})_{j} = \delta_{ij}$. 
Furthermore, 
there exists a unique stable Abelian differential $(\omega_{0})_{i}$ on $R_{0}$ such that 
its restriction $(\omega_{0})_{i}|_{\mathbb{CP}^{1}}$ to $\mathbb{CP}^{1}$ is $0$, 
$(\omega_{0})_{i}|_{E_{j}} = 0$ for $j \neq i$ and $(\omega_{0})_{i}|_{E_{i}}$ 
is the unique holomorphic Abelian differential on $E_{i}$ normalized for $A_{i}$. 
It is shown in \cite[(3.3)]{I4} (see also \cite{F}) that 
$$
(\omega_{s})_{i} \rightarrow (\omega_{0})_{i} \quad \mbox{under $s \rightarrow {\bf 0}$.} 
\eqno(2.4)
$$ 
Furthermore, it is shown in \cite[Section 3]{I4} that 
for $x \in E_{i} \setminus \{ t_{i} \}, y \in \mathbb{CP}^{1} \setminus \{ x_{1},..., x_{g} \}$, 
$$
\lim_{s \rightarrow {\bf 0}} (g_{R_{s}})_{j}^{k i_{1} \cdots i_{r}} (x, y) = 
\left\{ \begin{array}{ll} 
(g_{E_{i}})_{i}^{i i \cdots i}(x, t_{i}) & (\mbox{if $i = j = k = i_{1} = \cdots = i_{r}$}), 
\\ 
0 & (\mbox{otherwise}). 
\end{array} \right. 
$$
Therefore, 
if $x \in E_{i} \setminus \{ t_{i} \}, y \in \mathbb{CP}^{1} \setminus \{ x_{1},..., x_{g} \}$, 
then the connection form $\mathcal{K}_{R_{s}}(x, y)$ on $R_{s}$ given in (2.1) satisfies 
$$
\lim_{s \rightarrow {\bf 0}} \mathcal{K}_{R_{s}}(x, y; a, b) = \mathcal{K}_{E_{i}} (x, t_{i}; \pi_{i}(a), \pi_{i}(b)),  
\eqno(2.5) 
$$ 
where $\pi_{i}$ denotes the Lie algebra homomorphism defined on $\mathfrak{g}$ such that 
$\pi_{i}(a_{j}) = \delta_{ij} a_{i}$ and $\pi_{i}(b_{j}) = \delta_{ij} b_{i}$. 
As its corollary, 
$\lim_{s \rightarrow {\bf 0}} \mathcal{K}_{R_{s}}(x, y; a, b)$ has a simple pole at $x = t_{i}$ 
with residue $[b_{i}, a_{i}]$ (cf. \cite[Section 8]{E1} and \cite[Section 4.2]{DS1}). 
This implies that if $x, y \in \mathbb{CP}^{1} \setminus \{ x_{1},..., x_{g} \}$, 
then 
$$
\lim_{s \rightarrow {\bf 0}} \mathcal{K}_{R_{s}}(x, y; a, b) = 
- \sum_{i=1}^{g} [b_{i}, a_{i}]  \left( \frac{dx}{x - x_{i}} - \frac{dx}{x - y} \right) 
\eqno(2.6) 
$$
which is the KZ connection form on $\mathbb{CP}^{1}$ with trivial underlying bundle 
which has simple poles at $x = x_{i}$ with residues $-[b_{i}, a_{i}]$ $(i = 1,..., g)$, 
and $x = y$ with residue $\sum_{i=1}^{g} [b_{i}, a_{i}]$.

\section{The DHS connection and its variation} 

\subsection{The DHS connection}

The DHS connection form on a Riemann surface $R$ 
$$
\mathcal{J}_{R}(x, p; a, b) = \mathcal{J}^{\rm DHS}_{R}(x, p; a, b)
$$ 
is a single-valued smooth differential form 
in $x \in R \setminus \{ p \}$ which is the sum of a $(1, 0)$-form $\mathcal{J}^{(1,0)}_{R}$ 
which has a simple pole at $x = p$, 
and a $(0, 1)$-form $\mathcal{J}^{(0,1)}_{R}$ which is purely anti-holomorphic and 
single-valued on $R$. 
The connection form $\mathcal{J}_{R}$ takes values in the Lie algebra $\mathfrak{g}$ freely generated by $a_{i}, b_{i}$ $(i = 1,..., g)$. 
In \cite{DHS}, 
$\mathcal{J}_{R}$ was constructed in terms of higher convolutions on $R$ 
of the Arakelov Green functions and Abelian differentials, 
and shown to have the modular invariance. 
Furthermore, it was characterized in \cite[Theorem 2.5]{DESZ} 
by the following functional properties. 
For any fixed $p \in R$, 
there exists a unique differential form $\mathcal{J}_{R}(x, p; a, b)$ in $x$ which is 
smooth single-valued on $R \setminus \{ p \}$ and satisfies 
\begin{itemize}

\item 
the Maurer–Cartan equation for $x \neq p$, and it has a simple pole in $x = p$: 
$$
d_{x} \mathcal{J}_{R}(x, p; a, b) - \mathcal{J}_{R}(x, p; a, b) \wedge \mathcal{J}_{R}(x, p; a, b) 
= 2 \pi \sqrt{-1} \delta(x, p) \sum_{i=1}^{g} [b_{i}, a_{i}]. 
$$

\item 
its $(0, 1)$ component is given by 
$$
\mathcal{J}^{(0,1)}_{R}(x, p; a, b) = 
- \pi \sum_{i=1}^{g} b_{i} \left( {\rm Im}(\Omega)^{-1} \right)_{ij} \overline{\omega_{j}}(x); 
\quad \overline{\omega_{j}}(x) := \overline{\omega_{j}(x)}. 
\eqno(3.1) 
$$

\item
its $(1, 0)$ component $\mathcal{J}^{(1,0)}_{R}(x, p; a, b)$ is linear in the generators 
$a_{1},..., a_{g}$. 

\end{itemize}

Denote by $F(x, \alpha)$ the multi-valued function on the elliptic curve 
$E = \mathbb{C}/(\mathbb{Z} + \tau \mathbb{Z})$ defined in Section 2.1, 
and put 
$$
\Omega(x; \alpha) = 
\exp \left( 2 \pi \sqrt{-1} \frac{{\rm Im}(x)}{{\rm Im}(\tau)} \alpha \right) F(x; \alpha). 
$$
Then $\mathcal{J}_{E}$ is the Brown-Levin (BL) connection described as 
$$
\mathcal{J}_{E} (x, p; a, b) = dx \, \Omega(x - p; B) Ba - d \bar{x} \frac{\pi b}{{\rm Im}(\tau)}; 
\quad BX := [b, X] 
\eqno(3.2)
$$
which has a simple pole with residue $[b, a]$ at $x = p$, 
and degenerates to the KZ connection (2.2) on $\mathbb{CP}^{1}$ 
under ${\rm Im}(\tau) \rightarrow \infty$. 

The Brown-Levin (BL) elliptic polylogarithms \cite{BL} are obtained 
as monodromies of this DHS connection, 
and in \cite{BMMS, BDDT, BDDPT1, BDDPT2, BSZ1, BDDPT3, BK, MS1, MS2, W}, 
the elliptic multiple zeta values are defined as regularized iterated integrals of 
$\mathcal{J}_{E}$ from $x = 0$ to $1$ and to $\tau$, 
and applied to evaluating string amplitudes and Feynman integrals. 
In particular, see \cite{BSZ1} for the study of their $q$-expansion 
and relations to modular graph functions.

\subsection{Relation with the Enriquez connection}

We review results of \cite[Section 4]{DESZ} on the relationship between the DHS connection 
and the Enriquez connection. 
Following \cite[(4.2)]{DESZ}, 
denote by $\Gamma^{-}_{R}(x, y; b)$ the path-ordered exponential of $\mathcal{J}^{(0,1)}_{R}(t;b)$ 
from $y$ to $x$, 
namely 
$$
\Gamma^{-}_{R}(x, y; b) = {\rm P} \exp \int_{y}^{x} \mathcal{J}^{(0,1)}_{R}(t;b), 
$$
and as in \cite[(4.6)]{DESZ}, for $\gamma \in \pi_{1}(R)$, 
$$
\mu^{-}_{R}(\gamma, y; b) = \Gamma^{-}_{R}(\gamma \cdot y, y; b). 
$$
Furthermore, as in \cite[(4.10) and (4.12)]{DESZ}, put 
\begin{align*}
\Gamma^{\rm E}_{R}(x, y, p; \xi, \eta) 
& = {\rm P} \exp \int_{y}^{x} \mathcal{K}_{R}(t, p; \xi, \eta), 
\\  
\mu^{\rm E}_{R}(\gamma, y, p; \xi, \eta) 
& = \Gamma^{\rm E}_{R}(\gamma \cdot y, y, p; \xi, \eta). 
\end{align*}
Denote by $\mathfrak{g}_{b}$ the completed subalgebra of $\mathfrak{g}$ freely generated by $b_{1},..., b_{g}$. 
Then as is shown in \cite[Lemma 4.1 and (4.18)]{DESZ}, 
there exist uniquely vectors 
$$
\check{\xi}(y, p) = (\check{\xi}(y, p)_{1},..., \check{\xi}(y, p)_{g}), \quad 
\check{\eta}(y, p) = (\check{\eta}(y, p)_{1},..., \check{\eta}(y, p)_{g}) 
$$
of $(\mathfrak{g}_{b})^{g}$ such that 
\begin{align*}
\mu^{\rm E}_{R} \left( A_{k}, y, p; \check{\xi}(y, p), \check{\eta}(y, p) \right) 
& = \mu^{-}_{R} (A_{k}, y; b), 
\\ 
e^{2 \pi \sqrt{-1} \check{\eta}_{k}} 
\mu^{\rm E}_{R} \left( B_{k}, y, p; \check{\xi}(y, p), \check{\eta}(y, p) \right) 
& = \mu^{-}_{R} (B_{k}, y; b) 
\end{align*}
for all $k = 1,..., g$, 
and $\check{\xi}, \check{\eta}$ satisfy 
$$
(\check{\xi}_{1} \cdots \check{\xi}_{g})^{T} = 
- \pi {\rm Im}(\Omega)^{-1} (b_{1} \cdots b_{g})^{T} + O([b, b]), 
\quad \check{\eta}_{j} = b_{j} + O([b, b]). 
\eqno(3.3) 
$$
Furthermore, it is shown in \cite[Corollary 4.3]{DESZ} that 
$\mathcal{K}_{R}(t, p; \check{\xi}, \check{\eta})$ is independent of $p$, 
and hence as in \cite[Definition 4.6]{DESZ}, 
one can define 
$$
\Gamma^{\rm E}_{R}(x, p, \cdot \, ; \check{\xi}, \check{\eta}) 
= {\rm P} \exp \int_{p}^{x} \mathcal{K}_{R}(t, \cdot \, ; \check{\xi}, \check{\eta}). 
$$
Then by \cite[Corollary 4.9]{DESZ}, 
$$
\mathcal{U}^{\rm E}_{R}(x, p; \check{\xi}, \check{\eta}) = 
\Gamma^{\rm E}_{R}(x, p, \cdot \, ; \check{\xi}, \check{\eta}) \cdot \Gamma^{-}_{R}(x, p; b)^{-1}
$$
satisfies 
$$
\mathcal{J}^{(1,0)}_{R}(x, p; a, b) 
= \mathcal{U}^{\rm E}_{R}(x, p; \check{\xi}, \check{\eta})^{-1} \cdot 
\mathcal{K}_{R}(x, p; \tilde{a}, \check{\eta}) \cdot 
\mathcal{U}^{\rm E}_{R}(x, p; \check{\xi}, \check{\eta}), 
\eqno(3.4)
$$
where the elements $\tilde{a}_{1},..., \tilde{a}_{g}$ of $\mathfrak{g}$ are uniquely determined 
by the conditions that 
$$
\sum_{j=1}^{g} [b_{j}, a_{j}] = \sum_{j=1}^{g} [\check{\eta}_{j}, \tilde{a}_{j}]
$$ 
and that the replacement
$$
(a_{1},..., a_{g}, b_{1},..., b_{g}) \rightarrow (\tilde{a}_{1},..., \tilde{a}_{g}, \check{\eta}_{1},..., \check{\eta}_{g}) 
$$ 
gives an automorphism of the Lie algebra $\mathfrak{g}$ (cf. \cite[Lemma 4.7]{DESZ}).

\subsection{Variation of the DHS connection}

We show the variation of the DHS connection and its corollary similar to 
those for the Enriquez connection given in (2.5). 
Let $\{ R_{s} \}_{s}$ be the family of Riemann surfaces of genus $g$ as in Section 2.2 
which consists of deformations of 
$$
R_{0} = \mathbb{CP}^{1} \bigcup_{x_{i} = t_{i}} \left( \bigcup_{i=1}^{g} E_{i} \right) 
$$
by the relations (2.3). 
Then for $x \in E_{i} \setminus \{ t_{i} \}, p \in \mathbb{CP}^{1} \setminus \{ x_{1},..., x_{g} \}$, 
we will show that 
$$
\lim_{s \rightarrow {\bf 0}} \mathcal{J}_{R_{s}}(x, p; a, b) 
= \mathcal{J}_{E_{i}}(x, t_{i}; \pi_{i}(a), \pi_{i}(b)), 
$$ 
where $\pi_{i}$ denotes the Lie algebra homomorphism defined on $\mathfrak{g}$ such that 
$\pi_{i}(a_{j}) = \delta_{ij} a_{j}$ and $\pi_{i}(b_{j}) = \delta_{ij} b_{j}$. 
Let $\{ A_{i}, B_{i} \}_{i}$ be the symplectic homology basis of $R_{s}$ as in Section 2.2, 
and denote by $(\omega_{s})_{i}$ and $\Omega_{s}$ the associated normalized Abelian differentials 
and period matrix for $R_{s}$ respectively. 
Then as is seen in (2.4), 
$\lim_{s \rightarrow {\bf 0}} (\omega_{s})_{i} = (\omega_{0})_{i}$ is $0$ on $\mathbb{CP}^{1}$ 
and on $E_{j}$ for $j \neq i$, 
and it is the Abelian differential on $E_{i}$ normalized for $A_{i}$. 
Therefore, 
by putting $\tau_{j} = \oint_{B_{j}} (\omega_{0})_{j}$, 
$$
\lim_{s \rightarrow {\bf 0}} {\rm Im} (\Omega_{s}) 
= {\rm Im} \left( \oint_{B_{i}} (\omega_{0})_{j} \right)_{i,j} 
= \left( \delta_{ij} \cdot {\rm Im} (\tau_{j}) \right)_{i,j}, 
$$
and hence by (3.1), 
$$
\lim_{s \rightarrow {\bf 0}} \mathcal{J}^{(0,1)}_{R_{s}} (x; b) 
= - \pi \sum_{j=1}^{g} b_{j} \left( {\rm Im} (\tau_{j})^{-1} \overline{(\omega_{0})_{j}}(x) \right)_{j} 
= \mathcal{J}^{(0,1)}_{E_{i}} (x; b_{i}), 
\eqno(3.5)
$$ 
and $\lim_{s \rightarrow {\bf 0}} \mathcal{J}^{(0,1)}_{R_{s}} (t; b) = 0$ 
for $t \in \mathbb{CP}^{1} \setminus \{ x_{1},..., x_{g} \}$. 
Since each path from $p$ to $x$ is decomposed to two paths in $\mathbb{CP}^{1}$ 
(resp. $E_{i}$) from $p$ (resp. $t_{i}$) to $x_{i}$ (resp. $x$) under $s \rightarrow {\bf 0}$, 
by (3.5), 
$$
\lim_{s \rightarrow {\bf 0}} \Gamma^{-}_{R_{s}} (x, p; b) 
= {\rm P} \exp \int_{t_{i}}^{x} \mathcal{J}^{(0,1)}_{E_{i}}(t; b_{i}) 
= \Gamma^{-}_{E_{i}} (x, t_{i}; \pi_{i}(b)). 
\eqno(3.6)
$$ 
Furthermore, by (3.3), 
\begin{align*}
\lim_{s \rightarrow {\bf 0}} \pi_{j}(\check{\xi}_{k}) 
& = \frac{- \delta_{jk} \pi}{{\rm Im}(\tau_{j})} b_{j} + O([b_{j}, b_{j}]) 
= \frac{- \delta_{jk} \pi}{{\rm Im}(\tau_{j})} b_{j},  
\\ 
\pi_{j}(\check{\eta}_{k}) 
& = \delta_{jk} b_{j} + O([b_{j}, b_{j}]) = \delta_{jk} b_{j} 
\end{align*}
for any $j = 1,..., g$, 
and hence by (2.6), 
$$
\lim_{s \rightarrow {\bf 0}} \mathcal{K}_{R_{s}}(t, p; \check{\xi}, \check{\eta}) 
= - \sum_{i=1}^{g} \left[ b_{i}, - \frac{\pi}{{\rm Im}(\tau_{i})} b_{i} \right]  
\left( \frac{dt}{t - x_{i}} - \frac{dt}{t - p} \right) 
= 0 
\eqno(3.7) 
$$
for $t \in\mathbb{CP}^{1}$. 
Therefore, 
for $x \in E_{i} \setminus \{ t_{i} \}$ and 
$p \in \mathbb{CP}^{1} \setminus \{ x_{1},..., x_{g} \}$, 
by (3.7) and (2.5), 
\begin{align*}
\lim_{s \rightarrow {\bf 0}} \Gamma^{\rm E}_{R_{s}} (x, p; \check{\xi}, \check{\eta}) 
& 
= {\rm P} \exp \int_{p}^{x} \lim_{s \rightarrow {\bf 0}} 
\mathcal{K}_{R_{s}} (t, u; \check{\xi}, \check{\eta}) 
\quad (u \in \mathbb{CP}^{1} \setminus \{ x_{1},..., x_{g} \}) 
\\
& 
= {\rm P} \exp \int_{t_{i}}^{x} \mathcal{K}_{E_{i}} (t, t_{i}; \pi_{i}(\check{\xi}), \pi_{i}(\check{\eta})) 
\\
&
= \Gamma^{\rm E}_{E_{i}} (x, t_{i}; \pi_{i}(\check{\xi}), \pi_{i}(\check{\eta})) 
\end{align*}
which together with (3.6) imply 
\begin{align*}
\lim_{s \rightarrow {\bf 0}} \mathcal{U}^{\rm E}_{R_{s}} (x, p; \check{\xi}, \check{\eta})  
& = \lim_{s \rightarrow {\bf 0}} \Gamma^{\rm E}_{R_{s}} (x, p; \check{\xi}, \check{\eta}) \cdot 
\lim_{s \rightarrow {\bf 0}} \Gamma^{-}_{R_{s}} (x, p; b)^{-1}   
\\
& = \Gamma^{\rm E}_{E_{i}} (x, t_{i}; \pi_{i}(\check{\xi}), \pi_{i}(\check{\eta})) \cdot 
\Gamma^{-}_{E_{i}} (x, t_{i}; \pi_{i}(b))^{-1} 
\\
& = \mathcal{U}^{\rm E}_{E_{i}} (x, t_{i}; \pi_{i}(\check{\xi}), \pi_{i}(\check{\eta})). 
\end{align*}
Since $\pi_{i}(\check{\eta}_{j}) = \delta_{ij} b_{i}$ and hence $[b_{i}, a_{i}] = [b_{i}, \pi(\tilde{a}_{i})]$, 
by (3.4) and (2.5), 
\begin{eqnarray*}
\lefteqn{
\lim_{s \rightarrow {\bf 0}} \mathcal{J}^{(1,0)}_{R_{s}} (x, p; a, b) 
}
\\
& = & 
\mathcal{U}^{\rm E}_{E_{i}} (x, t_{i}; \pi_{i}(\check{\xi}), \pi_{i}(\check{\eta}))^{-1} \cdot 
\mathcal{K}_{E_{i}} (x, t_{i}; \pi_{i}(\tilde{a}), \pi_{i}(\check{\eta})) \cdot 
\mathcal{U}^{\rm E}_{E_{i}} (x, t_{i}; \pi_{i}(\check{\xi}), \pi_{i}(\check{\eta})) 
\\ 
& = & 
\mathcal{J}^{(1,0)}_{E_{i}} (x, t_{i}; a_{i}, b_{i}). 
\end{eqnarray*}
This together with (3.5) imply the main result of this section that 
$$
\lim_{s \rightarrow {\bf 0}} \mathcal{J}_{R_{s}} (x, p; a, b) = \mathcal{J}_{E_{i}} (x, t_{i}; a_{i}, b_{i}) 
$$
which is seen in Section 3.1 to have a simple pole with residue $[b_{i}, a_{i}]$ at $x = t_{i}$. 
Therefore, similarly as for the Enriquez connection, 
if $x, y \in \mathbb{CP}^{1} \setminus \{ x_{1},..., x_{g} \}$, 
then $\lim_{s \rightarrow {\bf 0}} \mathcal{J}_{R_{s}}(x, y)$ becomes the KZ connection form 
$\mathcal{K}_{\mathbb{CP}^{1}}$ on $\mathbb{CP}^{1}$ which has simple poles at $x = x_{i}$ 
with residues $-[b_{i}, a_{i}]$ $(i = 1,..., g)$, 
and $x = y$ with residue $\sum_{i=1}^{g} [b_{i}, a_{i}]$.

\section{The DHS higher genus polylogarithm}

\subsection{Family of pointed Riemann surfaces} 

We consider families of pointed Riemann surfaces which are deformations of 
a degenerate pointed complex curve with nodal singularities obtained as 
a union of Riemann spheres $\mathbb{CP}^{1}$. 
Let $\Delta = (V, E, T)$ be a connected graph which is a collection of 
three finite sets $V$ of vertices, $E$ of edges, $T$ of tails and two boundary maps 
$$
b : T \rightarrow V, \quad 
b : E \longrightarrow \left( V \cup \{ \mbox{unordered pairs of elements of $V$} \} \right). 
$$
We call $\Delta$ {\it of $(g, n)$-type} if $g = \dim H_{1}(\Delta)$ and $n$ is the number $|T|$ 
of elements in $T$, 
and take an {\it orientation} of $\Delta$, namely, an orientation of each $e$ in $E$. 
Under an orientation of $\Delta$, 
denote by $\pm E = \{ e, -e \ | \ e \in E \}$ the set of oriented edges, 
and by $v_{h}$ the terminal vertex of $h$ in $\pm E$ (resp. the boundary vertex of $h$ in $T$). 
For each $h$ in $\pm E$, 
denote by $|h|$ the corresponding element of $E$ without orientation. 

For an oriented graph $\Delta = (V, E, T)$ of $(g, n)$-type with orientation, 
take a subset $\mathcal{E}_{\infty}$ of $\pm E \cup T$ such that 
if $h, h' \in \mathcal{E}_{\infty}$ satisfy $v_{h} = v_{h'}$, 
then $h = h'$, 
and denote by $\mathcal{E}$ the complement of $\mathcal{E}_{\infty}$ 
in $\pm E \cup T$. 
We attach Riemann spheres $P_{v} = \mathbb{CP}^{1}$ for $v \in V$, 
and complex moduli parameters $x_{h}$ for $h$ in $\mathcal{E}$ such that 
$$
x_{h} \neq x_{h'} \ (\mbox{$h, h' \in \mathcal{E}$ with $h \neq h'$ and $v_{h} = v_{h'}$}).
$$
Then we obtain a degenerate complex curve $R_{\Delta}$ of genus $g$  
with $n$ marked points $x_{t}$ on $P_{v_{t}}$ for $t$ in $T$ 
by identifying $x_{h} \in P_{v_{h}}$ with $x_{-h} \in P_{v_{-h}}$ for $h$ in $\pm E$, 
where $x_{h} = \infty$ if $h$ in $\mathcal{E}_{\infty}$. 
For $h$ in $\pm E$, denote by $\xi_{h}$  denotes the local coordinate 
on $P_{v(h)} = \mathbb{CP}^{1}$ given by 
$$
\xi_{h}(z) = \left\{ \begin{array}{ll} 
z - x_{h} & \mbox{(if $x_{h} \neq \infty$),} 
\\
1/z   & \mbox{(if $x_{h} = \infty$).} 
\end{array} \right. 
$$
Then for non-zero complex deformation parameters $y_{e} = y_{-e}$ indexed by $e \in E$, 
$\mathcal{R}_{\Delta}$ (with marked points $x_{t}$ for $t$ in $T$) denotes the family of deformations of  $R_{\Delta}$ defined by gluing $\bigcup_{v \in V} P_{v}$ minus neighborhoods of $x_{h}$ via the relations 
$$
\xi_{h} \cdot \xi_{-h} = y_{h} \quad (h \in \pm E)
\eqno(4.1) 
$$
around the marked points $x_{\pm h} \in P_{v_{\pm h}}$ $(h \in \pm E)$. 

In what follows, we mainly consider the case when $\Delta = (V, E, T)$ is {\it trivalent} 
which means that each vertex in $V$ has just three branches. 
We fix a coordinate on each $P_{v} = \mathbb{CP}^{1}$ satisfying 
$$
\{ x_{h} \ | \ \mbox{$h \in \pm E \cup T$ with $v_{h} = v$} \} = \{ 0, 1, \infty \} 
\subset P_{v}. 
$$ 
Then $|E|$ is the dimension of the deformation space of $R_{\Delta}$ 
which is also the dimension $3g - 3 + n$ of the moduli space of Riemann surfaces 
of genus $g$ with $n$ marked points. 
Therefore, we have a fundamental relation 
$$
|E| = 3g - 3 + n. 
$$ 
Furthermore, $3 |V|$ is the sum of the number of tails and that of oriented edges, 
and hence 
$$
|V| = \frac{2 |E| + n}{3} = 2g - 2 + n. 
$$

\subsection{Gluing construction of the DHS connection} 

We construct the DHS connection on special families of $n$-pointed Riemann surfaces of genus $g$ 
by gluing the BL connections on elliptic curves and KZ connections on pointed Riemann spheres. 
Let 
$$
\Delta_{0} = (V_{0}, E_{0}, T_{0})
$$ 
be an oriented trivalent graph of $(g, n)$-type such that $E_{0}$ contains $g$ loops denoted by 
$\varepsilon_{1},..., \varepsilon_{g}$, 
namely $\Delta_{0}$ consists of its unique maximal subtree $M_{0}$ and $\varepsilon_{i}$ 
connected by vertices $v_{\varepsilon_{i}} = v_{-\varepsilon_{i}}$ $(i = 1,..., g)$. 
Then $\mathcal{R}_{\Delta_{0}}$ gives a family of deformations of the union $R_{\Delta_{0}}$ 
of $3$-pointed Riemann spheres $(P_{v}; x_{h}) = (\mathbb{CP}^{1}; 0, 1, \infty)$ 
for $v \in V_{0}$ and $h \in \pm E_{0} \cup T_{0}$ with $v_{h} = v$ by the relation (4.1) 
around $x_{\pm h} \in P_{v_{\pm h}}$ $(h \in \pm E_{0})$. 
For each $i = 1,..., g$, 
denote by $\mathcal{E}_{i}$ the elliptic curve associated with the loop $\varepsilon_{i}$ 
which is the deformation by $y_{\varepsilon_{i}}$ of the singular complex curve 
$$
P_{v_{\varepsilon_{i}}}/(x_{\varepsilon_{i}} = x_{\varepsilon_{-i}})
$$ 
obtained from $P_{v_{\varepsilon_{i}}} = P_{v_{-\varepsilon_{i}}}$ by identifying 
$x_{\varepsilon_{i}}$ and $x_{\varepsilon_{-i}}$ in $P_{v_{\varepsilon_{i}}}$. 
Then $\mathcal{R}_{\Delta_{0}}$ is also obtained from $P_{v}$ 
$(v \neq v_{\varepsilon_{1}},..., v_{\varepsilon_{g}})$ and $\mathcal{E}_{i}$ $(i = 1,..., g)$ 
by gluing around $x_{h}$ and $x_{-h}$ by the relation (4.1) 
for $h \neq \pm \varepsilon_{1},..., \pm \varepsilon_{g}$. 

Denote by $\tilde{\mathfrak{g}}$ the non-commutative algebra generated by 
the elements of $\mathfrak{g}$ and the symbols $X_{t}$ $(t \in T_{0})$ such that 
$$
\sum_{t \in T_{0}} X_{t} = \sum_{i=1}^{g} [b_{i}, a_{i}]. 
$$
Then by the induction on the numbers of edges in $M_{0}$, 
one can see that there exist uniquely elements $X_{h} \in \tilde{\mathfrak{g}}$, 
where $h$ are oriented edges or tails in $M_{0}$ satisfying the rules 
\begin{itemize}

\item 
For each $i = 1,..., g$, $X_{e_{i}} = [b_{i}, a_{i}]$, 
where $e_{i}$ denotes the unique oriented edge in $M_{0}$ such that $v_{e_{i}} = v_{\varepsilon_{i}}$. 

\item 
For each edge $e$ in $M_{0}$, $X_{e} + X_{-e} = 0$. 

\item 
For each vertex $v$ in $M_{0}$, $\sum_{h} X_{h} = 0$, 
where $h$ runs through the set of oriented edges or tails in $M_{0}$ such that $v_{h} = v$. 

\end{itemize} 
From the main result and its corollary in Section 3.3, 
we can construct the $\tilde{\mathfrak{g}}$-valued DHS connection on $\mathcal{R}_{\Delta_{0}}$ 
by gluing the KZ connections on $P_{v}$ $(v \neq v_{\varepsilon_{1}},..., v_{\varepsilon_{g}})$ and 
the BL connection on $\mathcal{E}_{i}$ $(i = 1,..., g)$ as follows. 
For each vertex  $v \neq v_{\varepsilon_{1}},..., v_{\varepsilon_{g}}$, 
denote by $\left( \mathcal{V}_{v}, \nabla_{v} \right)$ the KZ connection associated with 
$\sum_{v_{h} = v} d \xi_{h} / \xi_{h}$ on $P_{v} = \mathbb{CP}^{1}$, 
and for each $i = 1,..., g$, 
denote by $\left( \mathcal{V}_{i}, \nabla_{i} \right)$ the BL connection on $\mathcal{E}_{i}$ 
associated with $\mathcal{J}_{\mathcal{E}_{i}}(x, p; a_{i}, b_{i})$ given in (3.2) 
which has a simple pole at the marked point $p \neq x_{\pm \varepsilon_{i}}$ with residue $[b_{i}, a_{i}]$. 
Then by (4.1) and the above rule, 
$$
X_{h} \frac{d \xi_{h}}{\xi_{h}} = X_{-h} \frac{d \xi_{-h}}{\xi_{-h}} 
\quad \left( h \in \pm E_{0} \setminus \{ \pm \varepsilon_{1},..., \pm \varepsilon_{g} \} \right) 
$$ 
which implies that the monodromies 
$e^{2 \pi \sqrt{-1} X_{h}}$ of $X_{h} d \xi_{h}/\xi_{h}$ around $\xi_{h} = 0$ and 
$e^{2 \pi \sqrt{-1} X_{-h}}$ of $X_{-h} d \xi_{-h}/\xi_{-h}$ around $\xi_{-h} = 0$ satisfy 
$e^{2 \pi \sqrt{-1} X_{h}} \cdot e^{2 \pi \sqrt{-1} X_{-h}} = 1$. 
Therefore, 
one can glue $\left( \mathcal{V}_{v}, \nabla_{v} \right)$ 
$(v \neq v_{\varepsilon_{1}},..., v_{\varepsilon_{g}})$ and 
$\left( \mathcal{V}_{i}, \nabla_{i} \right)$ $(i = 1,..., g)$ around $y_{e} = 0$ 
$(e \neq \varepsilon_{1},..., \varepsilon_{g})$, 
and hence we obtain a vector bundle with flat connection 
$\left( \mathcal{V}_{\Delta_{0}}, \nabla_{\Delta_{0}} \right)$ on $\mathcal{R}_{\Delta_{0}}$ 
with simple poles at $x_{t}$ for $t$ in $T$. 
Since $\mathcal{R}_{\Delta_{0}}$ is a family of deformations of the singular complex curve  $R_{\Delta_{0}}$ by $y_{e}$ $(e \neq \varepsilon_{1},..., \varepsilon_{g})$, 
the monodromy of $\left( \mathcal{V}_{\Delta_{0}}, \nabla_{\Delta_{0}} \right)$ 
can be calculated combining the monodromies of the KZ connections 
$\left( \mathcal{V}_{v}, \nabla_{v} \right)$ and the BL connections 
$\left( \mathcal{V}_{i}, \nabla_{i} \right)$ together with the transition operators 
for the deformation. 
This calculation will be given in the following Section 4.3 
in the same way to \cite[Sections 2 and 3]{I3}.

\subsection{Calculation of the DHS polylogarithms} 

Let $\Delta_{0} = (V_{0}, E_{0}, T_{0})$ be the oriented trivalent graph of $(g, n)$-type 
given in Section 4.2 which contains the $g$ loops $\varepsilon_{1},..., \varepsilon_{g}$. 
Furthermore, 
we assume that $T_{0}$ is not empty. 

In the first step, 
we compute monodromies of the DHS connection described as 
$\left( \mathcal{V}_{\Delta_{0}}, \nabla_{\Delta_{0}} \right)$ 
on $\mathcal{R}_{\Delta_{0}} \setminus \{ x_{t} \ | \ t \in T_{0} \}$ 
between tangential points, namely, real and unit vectors at marked points $x_{t}$ $(t \in T_{0})$. 
These monodromies are expressed as products of the monodromies along 
\begin{itemize}

\item[(1)] 
paths between tangential points at $x_{h_{1}}, x_{h_{2}}$ on $P_{v}$ 
for $v \neq v_{\varepsilon_{1}},..., v_{\varepsilon_{g}}$ (see {\bf Figure 2}), 

\item[(2)] 
$A, B$-cycles in $\mathcal{E}_{i}$ $(i = 1,..., g)$ (see {\bf Figure 3}), 

\item[(3)] 
paths along $h$ from the tangential point associated with $\xi_{-h}$ to 
that associated with $\xi_{h}$ under the relation (4.1) 
for $h \neq \pm \varepsilon_{i},..., \pm \varepsilon_{g}$. 

\end{itemize}

\vspace{-5ex}
\hspace{-7ex} \includegraphics[width=150mm]{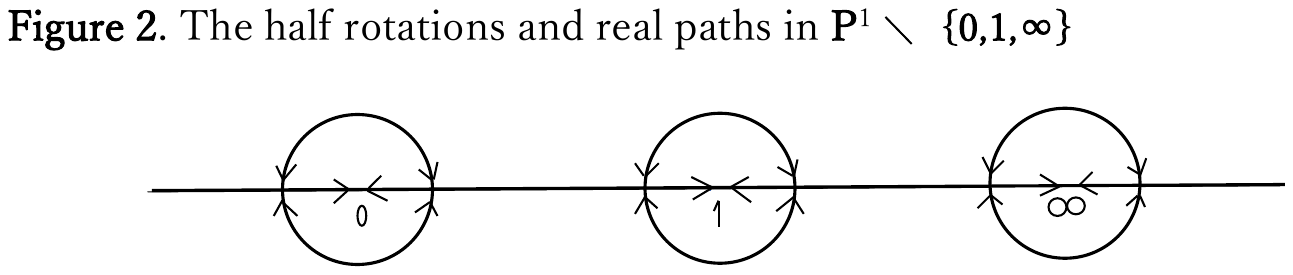}

\vspace{-45ex}

\hspace{-7ex} \includegraphics[width=150mm]{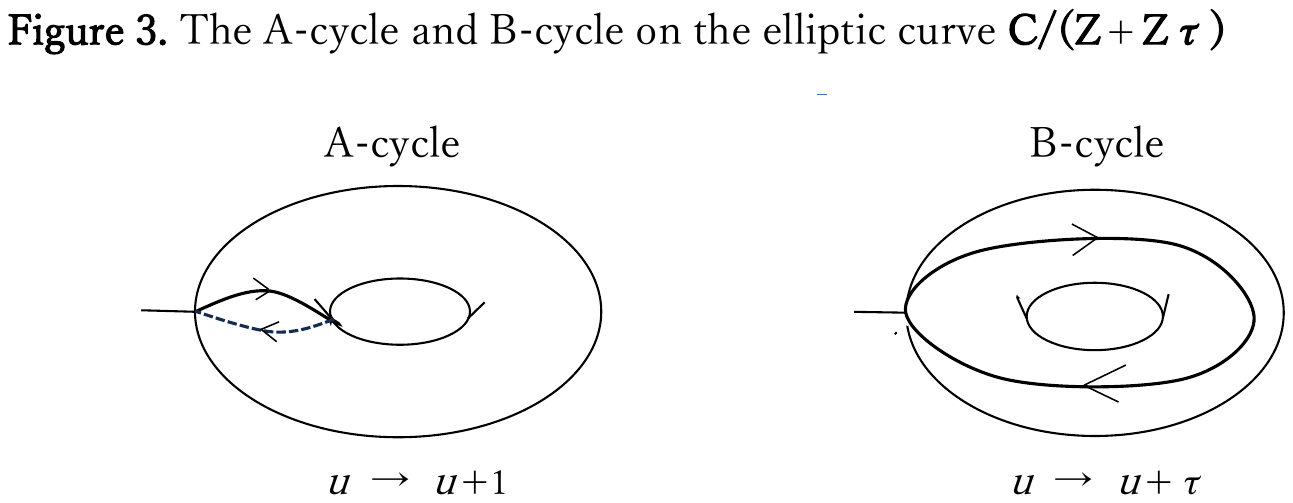}
\vspace{-28ex}

\noindent
We will explain the above monodromies more precisely. 
The monodromy along a path in (1) is the $\tilde{\mathfrak{g}}$-valued connection matrix 
for the flat sections $f_{i}(z)$ $(i = 1, 2)$ of the KZ connection $(\mathcal{V}_{v}, \nabla_{v})$ 
on $P_{v}$ which are normalized at $x_{h_{i}}$, namely 
$$
\lim_{\xi_{h_{i}}(z) \downarrow 0} \frac{f_{i}(z)}{\xi_{h_{i}}(z)^{X_{h_{i}}}} = 1. 
$$ 
Then this monodromy is expressed as a product of the {\it half-twist associators} 
$$
e^{\pm \pi \sqrt{-1} X_{h_{i}}} = 
\sum_{k=0}^{\infty} \frac{(\pi \sqrt{-1})^{k}}{k !} \left( \pm X_{h_{i}} \right)^{k}, 
$$ 
and the {\it KZ associators} 
$$
\Phi_{\rm KZ} \left( X_{h}, X_{h'} \right) 
\ \ (v_{h} = v_{h'} = v) 
$$
which are non-commutative formal power series in $X_{h}, X_{h'}$ 
whose coefficients are expressed by $\pi \sqrt{-1}$ and multiple zeta values. 
The monodromies along paths in (2) are the BL elliptic polylogarithms 
called the {\it A-cycle, B-cycle BL associators} respectively 
which are the $\tilde{\mathfrak{g}}$-valued connection matrix for the flat section $f(z)$ 
of the BL connection $(\mathcal{V}_{i}, \nabla_{i})$ on $\mathcal{E}_{i}$ which is normalized 
at $x_{\varepsilon_{i}}$ as 
$$
\lim_{\xi_{\varepsilon_{i}}(z) \downarrow 0} \frac{f(z)}{\xi_{\varepsilon_{i}}(z)^{[b_{i}, a_{i}]}} = 1. 
$$ 
As is seen in the last paragraph of Section 3.1, 
the coefficients of the BL associators are expressed by the elliptic multiple zeta values. 
Furthermore, the monodromy along the path in (3) is 
the $\tilde{\mathfrak{g}}$-valued connection matrix for the normalized flat sections 
$\xi_{h}^{X_{h}}$ and $\xi_{-h}^{X_{-h}}$ of $X_{h} d \xi_{h}/\xi_{h} = X_{-h} d \xi_{-h}/\xi_{-h}$ 
at $x_{h}$ and $x_{-h}$ respectively. 
Then by (4.1), this monodromy becomes 
$$
\left( \xi_{h}^{X_{h}} \right)^{-1} \cdot \xi_{-h}^{X_{-h}} = 
y_{h}^{X_{-h}} = \sum_{k=0}^{\infty} \frac{\log (y_{h})^{k}}{k!} (X_{-h})^{k} 
$$
which we call the {\it gluing associator} given by 
a multi-valued function of $y_{h}$ with values in $\tilde{\mathfrak{g}}$. 

We can calculate the products of these associators in the following way. 
For $i = 1, 2$, 
we consider the KZ connections on the $3$-pointed Riemann spheres 
$(P_{i} = \mathbb{CP}^{1}; 0, 1, \infty)$ with residues $(X_{i})_{0}$ at $0 \in P_{i}$ 
satisfying $(X_{1})_{0} + (X_{2})_{0} = 0$, 
and glue these connections around $0$ in $P_{1}$ and $P_{2}$ 
by the relation $z_{1} z_{2} = y$ for $z_{i} \in P_{i}$. 
Then the associated monodromy for the product of 
\begin{itemize}

\item 
the path $(1, 0)$ from the tangential point $\overrightarrow{10}$ at $1$ to 
that $\overrightarrow{01}$ at $0$ both in $P_{1}$, 

\item 
the path from the $\overrightarrow{01}$ in $P_{1}$ to $\overrightarrow{01}$ in $P_{2}$ 
by the above relation 

\end{itemize} 
is represented as the following product of the KZ and gluing associators
$$
z_{2}^{-(X_{2})_{0}} \cdot z_{1}^{(X_{1})_{0}} \cdot G_{0}(z)^{-1} \cdot G_{1}(z) = 
y^{(X_{1})_{0}} \cdot \Phi_{\rm KZ} \left( (X_{1})_{1}, (X_{1})_{0} \right),
$$
where $G_{0}(z), G_{1}(z)$ are the solutions of the KZ equation 
$$
d - \left( \frac{dz}{z} (X_{1})_{0} + \frac{dz}{z-1} (X_{1})_{1} \right) = 0
$$
on $P_{1}$ normalized at $\overrightarrow{0 1}, \overrightarrow{1 0}$ respectively. 

In the second step, 
we study the analytic continuation of the gluing associator $y_{h}^{X_{h}}$ 
along paths in (3) under a fusing (associative) move of $y_{h}$ 
from $\overrightarrow{01}$ to $\overrightarrow{10}$ by using a higher genus extension 
given in \cite[Appendix A]{I3} of the {\boldmath $u$}-equation 
$$
u_{ij} + \prod_{(k,l) \, {\rm crossing} \, (i,j)} u_{kl} = 1
$$
for the dihedral coordinates $u_{ij}$ in the genus $0$ case \cite{Br, AHHL, L}. 
This extension is based on the fact that cross-ratios of fixed points and ratios of eigenvalues 
for non-identity elements in Schottky groups give invariants of isomorphism classes 
of Riemann surfaces uniformized by these Schottky groups. 

\vspace{-5ex}
\hspace{-7ex} \includegraphics[width=150mm]{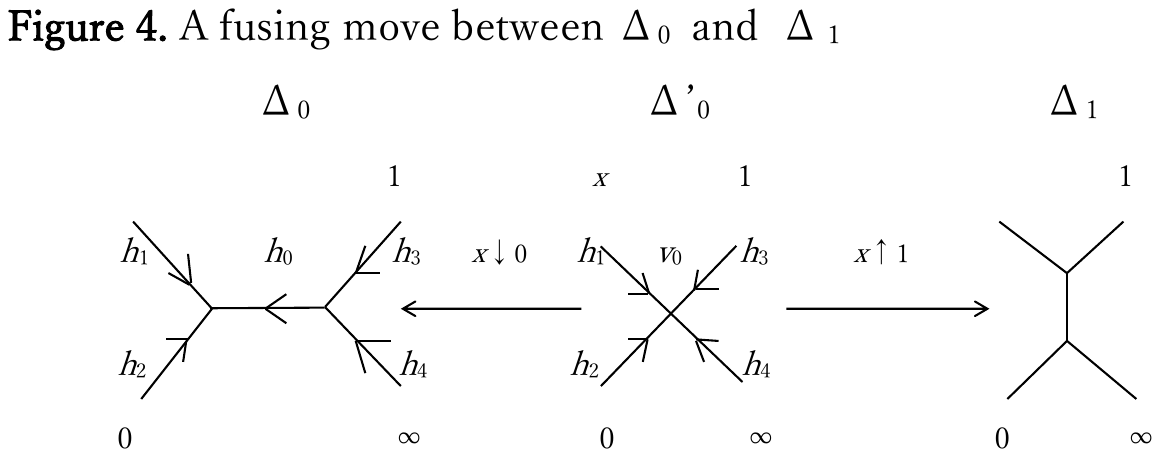}

\vspace{-28ex}

As in {\bf Figure 4}, 
we take a trivalent graph $\Delta_{1} = (V_{1}, E_{1}, T_{1})$ of $(g, n)$-type 
obtained from the above $\Delta_{0}$ by a fusing move for an oriented edge $h_{0}$ 
in $\pm E_{0} \setminus \{ \pm \varepsilon_{1},..., \pm \varepsilon_{g} \}$,  
and consider polylogarithms on the associated families $\mathcal{R}_{\Delta_{1}}$ 
of pointed complex curves. 
Denote by $\Delta'_{0} = \left( V'_{0} , E'_{0}, T'_{0} = T_{0} \right)$ the graph 
obtained from $\Delta_{0}$ by replacing $h_{0}$ with one vertex $v_{0}$. 
Then one can take $P_{v_{0}}$ as the Riemann sphere with four marked points 
$$
x_{h_{1}} = x, \quad x_{h_{2}} = 0, \quad x_{h_{3}} = 1, \quad x_{h_{4}} = \infty, 
$$
where $h_{i} \in \pm E'_{0}$ such that $v_{h_{i}} = v_{0}$, 
$x$ is approximated by the deformation parameter $y_{h_{0}}$ 
and $x \rightarrow 1$ gives $\mathcal{R}_{\Delta_{1}}$. 
We put $e_{i} = |h_{i}|$ $(i = 0,..., 3)$, 
and consider the family $\mathcal{R}_{\Delta'_{0}}$ as deformations of $R_{\Delta'_{0}}$ 
by parameters $y'_{e}$ $\left( e \in E'_{0} \right)$, 
namely $\mathcal{R}_{\Delta'_{0}}$ is obtained by $\xi_{h} \cdot \xi_{-h} = y'_{|h|}$ 
$\left( h \in \pm E'_{0} \right)$ around the marked points 
$x_{\pm h}$ on $P_{v_{\pm h}} = \mathbb{CP}^{1}$. 
We take parameters $t_{e}$ $\left( e \in E'_{0} \right)$ associated with the following cases: 
\vspace{2ex}

Case 1. If $h_{1} \neq -h_{2}, -h_{3}$, 
then we put 
$$
t_{e_{1}} = \frac{y'_{e_{1}}}{x(1-x)}, \quad t_{e_{2}} = \frac{y'_{e_{2}}}{x}, \quad 
t_{e_{3}} = \frac{y'_{e_{3}}}{1-x}, \quad t_{e} = y'_{e} \ 
\left( e \in E'_{0} \setminus \{ e_{1}, e_{2}, e_{3} \} \right).
$$

Case 2. If $h_{1} = -h_{2}$ (and then $h_{1} \neq -h_{3}$), 
then we put 
$$
t_{e_{1}} = \frac{y'_{e_{1}}}{1-x}, \quad t_{e_{2}} = y'_{e_{2}}, \quad 
t_{e_{3}} = \frac{y'_{e_{3}}}{1-x}, \quad t_{e} = y'_{e} \ 
\left( e \in E'_{0} \setminus \{ e_{1}, e_{2}, e_{3} \} \right).
$$

Case 3. If $h_{1} = -h_{3}$  (and then $h_{1} \neq -h_{2}$), 
then we put 
$$
t_{e_{1}} = \frac{y'_{e_{1}}}{x}, \quad t_{e_{2}} = \frac{y'_{e_{2}}}{x}, \quad 
t_{e_{3}} = y'_{e_{3}}, \quad t_{e} = y'_{e} \ 
\left( e \in E'_{0} \setminus \{ e_{1}, e_{2}, e_{3} \} \right).
$$

\noindent
Then as is shown in \cite[Appendix A]{I3}, 
in the above all cases, 
the following assertions hold: 
\begin{itemize}

\item
By moving $x$ and $t_{e}$ $\left( e \in E'_{0} \right)$ around $x = 0$, $t_{e} = 0$, 
we have deformations of $R_{\Delta_{0}}$. 

\item 
By moving $1- x$ and $t_{e}$ $\left( e \in E'_{0} \right)$ around $x = 1$, $t_{e} = 0$, 
we have deformations of $R_{\Delta_{1}}$. 

\item
The parameter $y_{e_{0}}$ is expressed as $y_{e_{0}} = x (1 - \alpha)$, 
where $\alpha$ is a computable power series in $x$ and $t_{e}$ 
which becomes $0$ when we put $t_{e_{i}} = 0$ $(i = 1,..., 4)$. 

\end{itemize}
Therefore, 
by expressing the above associators as $\tilde{\mathfrak{g}}$-valued functions of 
$x$, $t_{e}$ $\left( e \in E'_{0} \right)$ around $x = 0$, $t_{e} = 0$, 
and by moving $x$ from $0$ to $1$, 
we have the analytic continuation of the associators defined on $\mathcal{R}_{\Delta_{0}}$ 
to those defined on $\mathcal{R}_{\Delta_{1}}$. 
Especially, the gluing associators 
$$
y_{e_{0}}^{\pm X_{e_{0}}} = 
\sum_{k=0}^{\infty} \frac{\log (y_{e_{0}})^{k}}{k!} (\pm X_{e_{0}})^{k} 
\ \ \mbox{around $x = 0$, $t_{e} = 0$ $\left( e \in E'_{0} \right)$} 
$$
can be analytically continued to 
\begin{eqnarray*}
\lefteqn{
\sum_{k=0}^{\infty} 
\frac{\left[ \log (1 - (1-x)) + \log(1 - \alpha) \right]^{k}}{k!} (\pm X_{e_{0}})^{k} 
}
\\
& = & 
\sum_{k=0}^{\infty} \frac{1}{k!} 
\left( \sum_{l=1}^{\infty} \frac{(1-x)^{l} + \alpha^{l}}{l} \right)^{k} (\pm X_{e_{0}})^{k} 
\end{eqnarray*}
around $x = 1$, $t_{e} = 0$. 
This implies also that we can calculate the polylogarithms associated with 
$\left( \mathcal{V}_{\Delta_{0}}, \nabla_{\Delta_{0}} \right)$ on $\mathcal{R}_{\Delta_{1}}$. 

In the third step, 
by repeating the consideration in the second step concerned with a fusing move 
$$
\Delta_{0} = (V_{0}, E_{0}, T_{0}) \rightarrow \Delta_{1} = (V_{1}, E_{1}, T_{1}), 
$$
we calculate polylogarithms for degenerating families of Riemann surfaces 
of $(g, n)$-type. 
Denote by $\varepsilon_{1}^{(1)},..., \varepsilon_{g}^{(1)}$ the edges in $E_{1}$ 
corresponding to the loops $\varepsilon_{1},..., \varepsilon_{g}$ respectively 
under the inclusion $E_{0} \setminus \{ e_{0} \} \hookrightarrow E_{1}$. 
Note that $\varepsilon_{i}^{(1)}$ are possibly not loops. 
Then for integers $k > 1$,  
we take inductively trivalent graphs
$$
\Delta_{k} = (V_{k}, E_{k}, T_{k}) \ \ \mbox{with} \ \ 
\varepsilon_{1}^{(k)},..., \varepsilon_{g}^{(k)} \in E_{k}
$$ 
obtained from 
$$
\Delta_{k-1} = (V_{k-1}, E_{k-1}, T_{k-1}) \ \ \mbox{with} \ \ 
\varepsilon_{1}^{(k-1)},..., \varepsilon_{g}^{(k-1)} \in E_{k-1}
$$ 
by a fusing move on an edge in 
$E_{k-1} \setminus \left\{ \left. \varepsilon_{i}^{(k-1)} \ \right| i =1,..., g \right\}$ 
such that $\varepsilon_{i}^{(k)}$ are corresponding to $\varepsilon_{i}^{(k-1)}$. 
Then as in the second step, 
we can calculate the polylogarithms on $\mathcal{R}_{\Delta_{k}}$ 
by the analytic continuation. 
The fusing moves $\Delta_{k-1} \rightarrow \Delta_{k}$ correspond to fusing moves 
on the trivalent graphs of $(0, n + 2g)$-type obtained from $\Delta_{k-1}$ 
by cutting each $\varepsilon_{i}^{(k-1)}$ into two tails. 
Since all trivalent graphs of fixed type are connected by fusing moves (cf. \cite[Chapter 5]{BaK}), 
we can calculate explicitly the polylogarithms on families of Riemann surfaces of $(g, n)$-type 
which are close to maximally degenerate complex curves.

\section{Conclusion and outlook}
We studied the variation of the DHS connection and the associated higher genus polylogarithms 
under separating degenerations of Riemann surfaces with marked points. 
We showed that this DHS connection can be constructed by gluing the BL connection 
on elliptic curves and the KZ connections on pointed Riemann spheres. 
Therefore, 
one can describe explicitly the DHS higher genus polylogarithms defined on 
families of pointed Riemann surfaces in terms of multiple zeta values and 
elliptic multiple zeta values via deformation parameters with these families. 

For the theoretical study of the DHS connection, 
by the expression in 3.1 of the degenerated BL connection as the KZ connection, 
and the variation in 3.3 of the DHS connection, 
it is expected to describe the behavior of the DHS connection under separating and non-separating 
degenerations of pointed Riemann surfaces. 
See results of D'Hoker-Hidding-Schlotterer \cite[Theorem 6.3]{DHS} for the separating case 
in the genus two, 
and of Biancotto-D'Hoker-Kleinschmidt-Santagata-Schlotterer \cite{BDKSS} 
for the non-separating case. 
For the practical study of the DHS polylogarithms, 
we hope that our explicit formulas of the higher genus polylogarithms will be applicable to 
computing multi-loop Feynman integrals and string amplitudes as is done for the one-loop case 
in \cite{BMMS, BDDT, BDDPT1, BDDPT2, BSZ1, BDDPT3, BK, MS1, MS2}. 
\vspace{4ex} 

\noindent 
{\bf Declaration of competing interest} 
\vspace{1ex}

The author declares that he has no known competing financial interests 
or personal relationships that could have appeared to influence the work 
reported in this article. 
\vspace{2ex}

\noindent 
{\bf Data availability} 
\vspace{1ex}

No data was used for the research described in this article. 
\vspace{2ex}

\noindent 
{\bf Acknowledgments} 
\vspace{1ex}

This work is partially supported by the JSPS Grant-in-Aid for 
Scientific Research No. 25K06920.





\end{document}